\documentclass[fleqn,usenatbib]{mnras}

\usepackage{newtxtext,newtxmath}

\usepackage[T1]{fontenc}

\DeclareRobustCommand{\VAN}[3]{#2}
\let\VANthebibliography\thebibliography
\def\thebibliography{\DeclareRobustCommand{\VAN}[3]{##3}\VANthebibliography}

\usepackage{graphicx}	
\usepackage{amsmath}	
\usepackage{amssymb}	
\usepackage{lineno}
\usepackage[utf8]{inputenc}
\usepackage{color}
\usepackage{graphicx}
\usepackage{float}
\usepackage{acronym}
\usepackage{multirow}
\usepackage{tikz}
\usetikzlibrary{arrows,shapes,trees,decorations.pathreplacing}
\usepackage{import}
\usepackage{svn-multi}
\usepackage{lineno}
\usepackage{hyperref}
\usepackage{gensymb}
\hypersetup{
    colorlinks=true,
    linkcolor=blue,
    filecolor=magenta,      
    urlcolor=cyan,
}

\newcommand{\ecc}{\ensuremath{e}}

\newcommand{\chieff}{\ensuremath{\chi_{\mathrm{eff}}}}

\newcommand{\msun}{\ensuremath{\mathrm{M}_{\odot}}}
\newcommand{\chirpm}{\ensuremath{\mathcal{M}}}

\title[Eccentricity of GW170817 and GW190425]{Measuring the eccentricity of GW170817 and GW190425}

\author[A. K. Lenon et al.]{
Amber K. Lenon,$^{1}$\thanks{Email: alenon@syr.edu}
Alexander H. Nitz,$^{2,3}$
Duncan A. Brown$^{1,4}$
\\
$^{1}$Department of Physics, Syracuse University, Syracuse NY 13244, USA\\
$^{2}$Max-Planck-Institut f{\"u}r Gravitationsphysik (Albert-Einstein-Institut), D-30167 Hannover, Germany\\
$^{3}$Leibniz Universit{\"a}t Hannover, D-30167 Hannover, Germany\\
$^{4}$Kavli Institute for Theoretical Physics, University of California, Santa Barbara, CA 93106, USA
}

\date{Accepted XXX. Received YYY; in original form ZZZ}

\pubyear{2020}
\begin{document}
\label{firstpage}
\pagerange{\pageref{firstpage}--\pageref{lastpage}}
\maketitle

\begin{abstract}
Two binary neutron star mergers, GW170817 and GW190425, have been detected by Advanced LIGO and Virgo. These signals were detected by matched-filter searches that assume the star's orbit has circularized by the time their gravitational-wave emission is observable. This suggests that their eccentricity is low, but full parameter estimation of their eccentricity has not yet been performed. We use gravitational-wave observations to measure the eccentricity of GW170817 and GW190425. We find that the eccentricity at a gravitational-wave frequency of 10 Hz is  $e \leq 0.024$ and $e \leq 0.048$ for GW170817 and GW190425, respectively (90\% confidence). This is consistent with the binaries being formed in the field, as such systems are expected to have circularized to $e \leq 10^{-4}$ by the time they reach the LIGO-Virgo band. Our constraint is a factor of two smaller that an estimate based on GW170817 being detected by searches that neglect eccentricity. However, we caution that we find significant prior dependence in our limits, suggesting that there is limited information in the signals. We note that other techniques used to constrain binary neutron star eccentricity without full parameter estimation may miss degeneracies in the waveform, and that for future signals it will be important to perform full parameter estimation with accurate waveform templates.
\end{abstract}




\section{Introduction}
\label{sec:intro}
The Advanced LIGO and Virgo observatories have detected two binary neutron star mergers, GW170817 \citep{TheLIGOScientific:2017qsa} and GW190425 \citep{Abbott:2020uma}. To date, 17 double neutron star systems have been observed through radio surveys of the Milky Way field \citep{Martinez:2017jbp,Tauris:2017omb,Cameron:2017ody,Stovall:2018ouw,Lynch:2018zxo}. Observations of binary neutron stars allow us to determine their formation channels \citep{Smarr1976,Canal:1990dz,PortegiesZwart1:1997zn,Postnov:2006hka,Kalogera:2006uj,Kowalska:2010qg,Beniamini:2015uta,Tauris:2017omb,Palmese:2017yhz,Belczynski:2018ptv,Vigna-Gomez:2018dza,Giacobbo:2018etu,Mapelli:2018wys,Andrews:2019vou}, constrain the neutron-matter equation of state \citep{Bauswein:2017vtn,Annala:2017llu,Fattoyev:2017jql,De:2018uhw,Abbott:2018exr,Capano:2019eae,Tews:2018iwm,Most:2018hfd,Radice:2018ozg,Coughlin:2018fis,Forbes:2019xaz}, and test the strong-field regime of general relativity \citep{Abbott:2018lct}.

Although the eccentricity of double neutron stars in the Milky Way field ranges from $0.06$ to $0.828$ \citep{Zhu:2017znf,Andrews:2019vou}, field binaries will circularize to eccentricity $e \leq 10^{-4}$ \citep{Peters:1964zz,Kowalska:2010qg}, making them detectable by matched-filter searches that neglect eccentricity \citep{Martel:1999tm,Cokelaer:2009hj,Brown:2009ng,Huerta:2013qb}.
GW170817 and GW190425 were detected by searches that neglect eccentricity \citep{TheLIGOScientific:2017qsa,Abbott:2020uma}, suggesting that their eccentricity is $e \lesssim 0.05$ \citep{Huerta:2013qb}, however no full parameter estimation of their eccentricity of their eccentricity has been performed. The eccentricity of binary black hole observations \citep{Abbott:2016blz,Abbott:2016nmj,Abbott:2017vtc,Abbott:2017oio,Abbott:2017gyy,LIGOScientific:2020stg} has been explored in \cite{Romero-Shaw:2019itr} and \cite{Wu:2020zwr}. \cite{10.1093/mnrasl/slaa084} place a limit on the eccentricity of GW190425 by estimating the effect of eccentricity on the measured parameters of the signal. Here, we directly measure the eccentricity of GW170817 and GW190425 using Bayesian parameter estimation \citep{Biwer:2018osg}.

We use the observations from the Gravitational-Wave Open Science Center \citep{TheLIGOScientific:2017qsa,Abbott:2020uma}, waveform templates that include eccentricity \citep{Moore:2016qxz}, and Markov Chain Monte Carlo parameter estimation \citep{ForemanMackey:2012ig,Biwer:2018osg} to measure the eccentricity of the GW170817 and GW190425 when they have a gravitational-wave frequency of 10 Hz. We find that the eccentricity of GW170817 is $e \leq 0.024$ and GW190425 is $e \leq 0.048$ at 90\% confidence for a uniform prior on $e$. Our limit on eccentricity of GW170817 is a factor of two smaller than the limit estimated by its detection with circular waveform templates. We note that when using a common prior on eccentricity, our limit on the eccentricity of GW190425 is a factor of three greater than the limit of \cite{10.1093/mnrasl/slaa084}. This is due to a degeneracy between the chirp mass and eccentricity that is not included in the analysis of \cite{10.1093/mnrasl/slaa084}. However, this difference does not invalidate their conclusions about the formation of GW190425. When considering the two priors we use in our analysis, the limit on the eccentricity of GW190425 changes by a factor of two, suggesting that the eccentricity-constraining information in the signal is limited.

Dynamical interations may form binary neutron stars with residual eccentricity, although the rate of such mergers is expected to be small in current detectors \citep{Lee:2009ca,Ye:2019xvf} and a search for eccentric binary neutron stars in the O1 and O2 observing runs did not yield any candidates \citep{Nitz:2019spj}. However, since eccentricity is an interesting probe of binary formation channels and eccentric binaries may produce different electromagnetic emission than circular binary neutron stars \citep{Radice:2016dwd,Chaurasia:2018zhg}, it is important to accurately constrain the eccentricity of binary neutron stars as the number of observed mergers increases in the coming years.

\section{Methods}
\label{sec:method}

We measure the parameters of GW170817 and GW190425 using Bayseian inference \citep{Finn:2000hj,Rover:2006ni}. We use gravitational-wave data from Advanced LIGO and Virgo \citep{Blackburn:170817,dataLIGO:190425}, $\boldsymbol{d}(t)$, and a model of the gravitational waves, $H$, to calculate the posterior probability density function, $p(\boldsymbol{\theta}|\boldsymbol{d}(t),H)$, given by
\begin{equation}
    p(\boldsymbol{\theta}|\boldsymbol{d}(t),H) = \frac{p(\boldsymbol{\theta}|H) p(\boldsymbol{d}(t)|\boldsymbol{\theta},H)}{p(\boldsymbol{d}(t)|H)},
\end{equation}
where $\boldsymbol{\theta}$ denotes the parameters of the gravitational waveform, $p(\boldsymbol{\theta}|H)$, is the prior distribution on the signal parameters, and $p(\boldsymbol{d}(t)|\boldsymbol{\theta},H)$, is the probability of observing the data, known as the likelihood. The likelihood models the noise in the detector as a Gaussian and depends upon a noise-weighted inner product between the gravitational waveform and gravitational-wave data, $\boldsymbol{d}(t)$. Markov Chain Monte Carlo (MCMC) techniques can be used to marginalize over the parameters to obtain the posterior probabilities \citep{Christensen:2001cr}. Our implementation of Bayesian inferences uses the \textit{PyCBC Inference} software package \citep{Biwer:2018osg,alex_nitz_2020_3630601} and the parallel-tempered \textit{emcee} sampler, \texttt{emcee\_pt} \citep{ForemanMackey:2012ig,emceept}.

For GW170817 and GW190425, the MCMC is performed over the component masses of the binary, $m_{1,2}$, the component spins aligned with the orbital angular momentum, $\chi_{1,2}$, the time of coalescence, $t_c$, the polarization of the GW, $\psi$, the inclination angle, $\iota$, and the eccentricity, $\ecc$.

\begin{table*}
\begin{center}
\caption{Prior distributions and GPS time intervals for GW170817 and GW190425.}
    \begin{tabular}{|c||c|c|}
        \hline
        \textbf{Parameters}              & \textbf{GW170817}                 & \textbf{GW190425}                     \\ \hline
        \textbf{Component Masses} \msun  & {[}1.0,3.0{]}                     & {[}1.0,3.0{]}                         \\ \hline
        \textbf{Component Spins}         & {[}-0.05,0.05{]}                  & {[}-0.05,0.05{]}                      \\ \hline
        \textbf{Coalescence Time} (s)    & {[}1187008882.33,1187008882.53{]} & {[}1240215502.917,1240215503.117{]}   \\ \hline
        \textbf{Polarization}            & {[}0,2$\pi${]}                    & {[}0,2$\pi${]}                        \\ \hline
        \textbf{Inclination Angle}       & $\sin \iota$                      &  $\sin \iota$                         \\ \hline
        \textbf{Distance} (Mpc)          & 40.7 $\pm$ 2.36~\citep{Cantiello:2018ffy} & uniform in comoving volume                     \\ \hline
        \textbf{RA/Dec} ($\degree$)        & 3.44615914, -0.40808407~\citep{Soares-Santos:2017lru} & {[}$-\pi/2$,$\pi/2${]}           \\ \hline
        \textbf{Eccentricity}            & {[}0.0,0.1{]}                     & {[}0.0,0.1{]}                        \\ \hline 
        \hline
        \textbf{PSD Estimation Interval} (s) & {[}1187008382,1187008918{]}   & {[}1240215003,1240215543{]}          \\ \hline
        \textbf{Likelihood Interval} (s) & {[}1187008692,1187008892{]}       & {[}1240215313,1240215513{]}          \\ \hline
    \end{tabular}
\end{center}
\label{tab:prior}
\end{table*}

We assume a uniform prior distribution on the component masses, component spins, and coalescence time around the trigger shown in Table~\ref{tab:prior}. We assume an isotropic sky location for GW190425 and a prior uniform in $\sin \iota$ for the inclination angle of both detections. We fix the sky location of GW170817 to the observed EM counterpart using a Gaussian prior distribution on the distance \citep{Cantiello:2018ffy}. We explore the prior distribution on the eccentricity by running the MCMC with two prior distributions: a prior that is uniform in $\ecc$ and a prior uniform in $\log e$ to compare with the GW190425 results found by \cite{10.1093/mnrasl/slaa084}.

We use the GW strain data from the Advanced LIGO and Virgo detectors for GW170817 and GW190425, available through the LIGO Open Science Center (LOSC) \citep{Vallisneri:2014vxa}. The \texttt{LOSC\_CLN\_4\_V1} data that we use for GW170817 includes post-processing noise-subtraction performed by the LIGO/Virgo Collaboration \citep{Blackburn:170817,Driggers:2018gii}. The \texttt{T1700406\_v3} data that we use for GW190425 includes pre-processing glitch removal performed by the LIGO/Virgo Collaboration specifically for use in parameter estimation \citep{Abbott:2020uma,dataLIGO:190425}.

We high-pass the data using an eighth-order Butterworth filter with an attenuation of 0.1 at 15 Hz. To conserve the phase of the delay, the filter is applied forward and backwards. A low-pass finite impulse response filter is applied to the data prior to resampling. The data is decimated to 2048 Hz for the analysis. For computing the likelihood, we use Welch's method to estimate the detector's noise power spectral density (PSD). Welch's method is used with 16 second Hanning windowed segments that are overlapped by 8 seconds. The PSD is shortened to 8 seconds in the time domain \citep{Allen:2005fk}. The gravitational-wave data, $\boldsymbol{d}(t)$, used in the likelihood is taken from the intervals shown in Table~\ref{tab:prior}. The gravitational-wave likelihood is evaluated from a low-frequency cutoff of 20 Hz to the Nyquist Frequency of 1024 Hz.

A variety of waveforms are available that model eccentricity \citep{Huerta:2014eca,Tanay:2016zog,Moore:2016qxz,Huerta:2016rwp,Cao:2017ndf,Hinder:2017sxy,Tiwari:2019jtz,Moore:2019xkm}. From what we know of binary neutron star mergers, we expect them to have low mass, spin, and eccentricity making TaylorF2Ecc a suitable waveform. The waveform model, H, is TaylorF2Ecc, a TaylorF2 post-Newtonian (pN) model with eccentric corrections. We use the LIGO Algorithm Library implementation \citep{lalsuite} accurate to 3.5 pN order in orbital phase \citep{Buonanno:2009zt}, 3.5 pN order in the spin-orbit interactions \citep{Bohe:2013cla}, 2.0 pN order in spin-spin, quadrupole-monopole, and self-interactions of individual spins \citep{Mikoczi:2005dn,Arun:2008kb}, and 3.0 pN order in eccentricity \citep{Moore:2016qxz}. Since TaylorF2Ecc follows TaylorF2 in its construction, the waveform will terminate at twice the orbital frequency of a particle at the innermost stable circular orbit of a Schwarzschild black hole.

As a check on our analysis, we estimate the parameters of GW170817 and GW190425 using two available waveforms: the TaylorF2Ecc waveform at $e=0$ and the TaylorF2 waveform. Our analyses are consistent with each other and with the parameters estimated by Advanced LIGO and Virgo \citep{TheLIGOScientific:2017qsa,Abbott:2020uma}. 

\section{Results}
\label{sec:Results}
\begin{figure*}
    \includegraphics[width=1.0\textwidth]{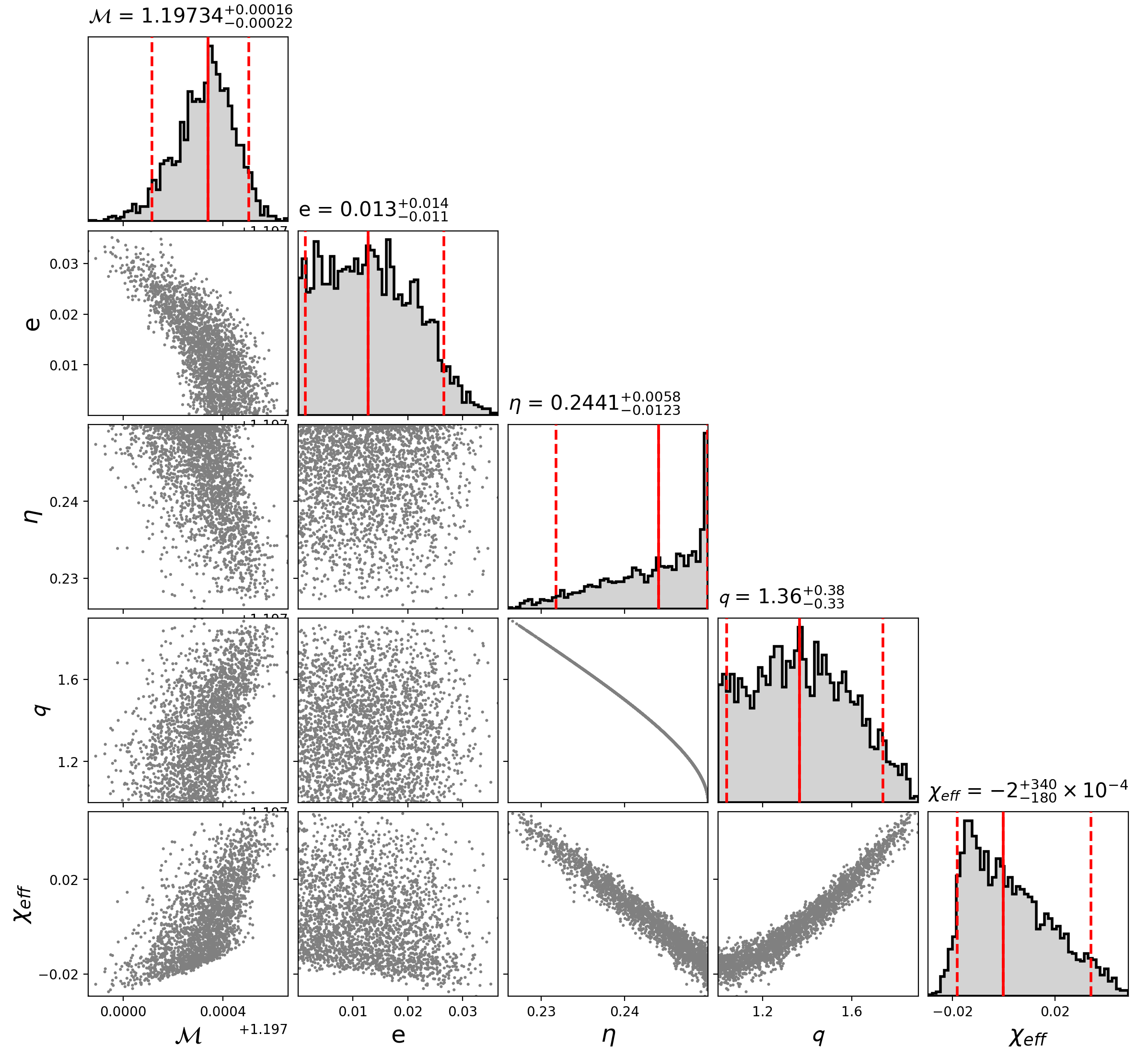}
    \caption{Posterior probability distribution of GW170817 at 10 Hz. The analysis used a prior uniform in $\ecc$. Each parameter is quoted with a median value (solid red line) and a 90\% credible interval (dashed red lines). The chirp mass $\chirpm$ is given in the detector frame. Note the degeneracy between $\chirpm$ and $\ecc$.}
\label{Fig:GW170817}
\end{figure*}

\begin{figure*}
    \includegraphics[width=1.0\textwidth]{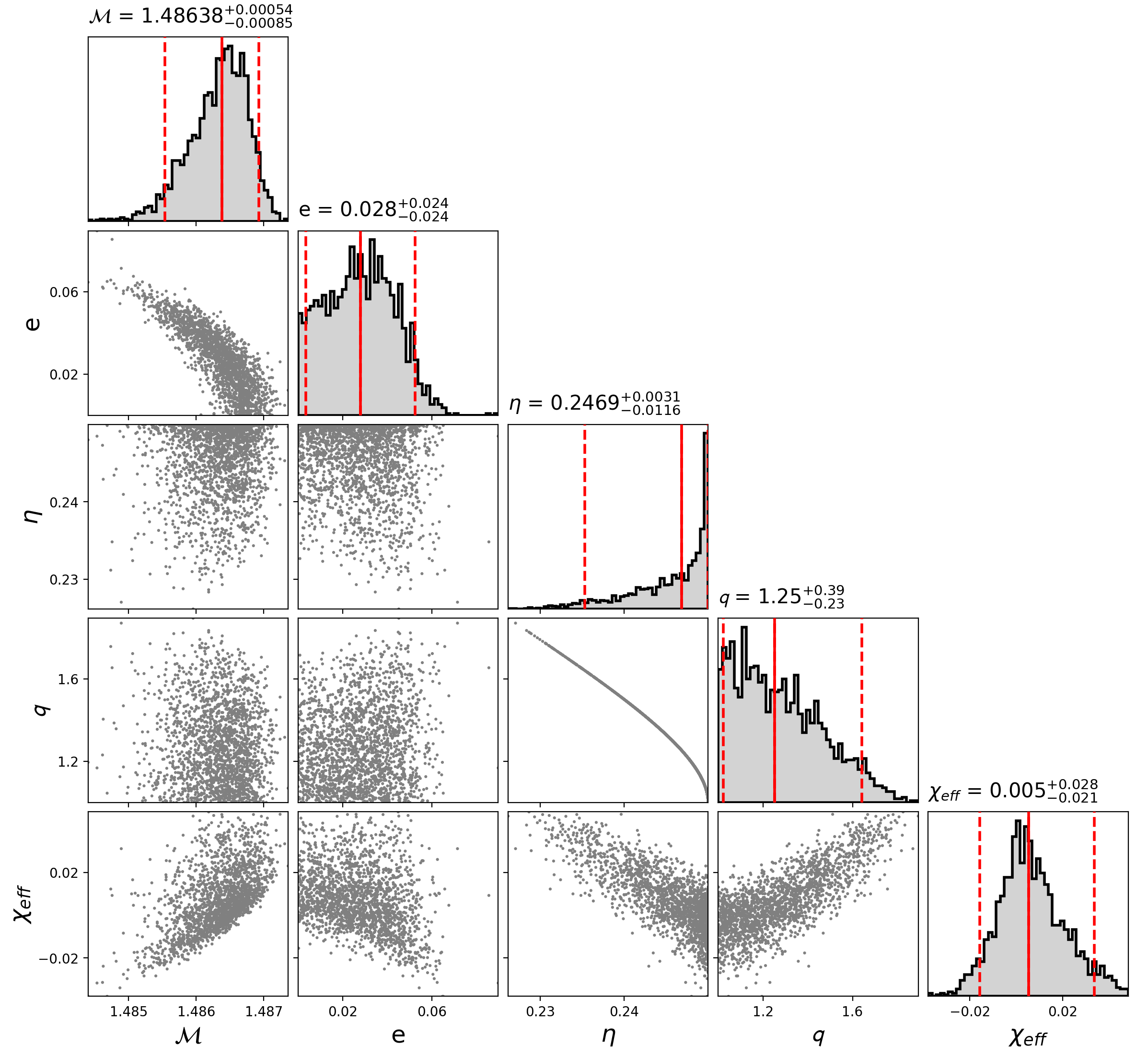}
    \caption{Posterior probability distribution of GW190425 at 10 Hz. The analysis used a prior uniform in $\ecc$. Each parameter is quoted with a median value (solid red line) and a 90\% credible interval (dashed red lines). The chirp mass $\mathcal{M}$ is given in the detector frame. Note the degeneracy between $\chirpm$ and $\ecc$.}
\label{Fig:GW190425}
\end{figure*}

We first constrain the level of the eccentricity by using the TaylorF2Ecc waveform and a prior uniform in $\ecc$. We find that the 90\% credible intervals at 10 Hz for GW170817 and GW190425 are $e = 0.012^{+0.013}_{-0.012}$ and $e = 0.025^{+0.022}_{-0.025}$ respectively. A degeneracy between the chirp mass, $\chirpm$, and eccentricity, $\ecc$ and a small correlation between the effective spin, $\chieff$, and $\ecc$ are shown in our posterior distributions in Figure~\ref{Fig:GW170817} and Figure~\ref{Fig:GW190425}. Since $\chirpm$ and $\chieff$ are correlated \citep{Baird:2012cu,Safarzadeh:2020mlb}, this will create a small correlation between $\ecc$ and $\chieff$.

\cite{10.1093/mnrasl/slaa084} estimated the eccentricity of GW190425 to determine if the formation channel was due to unstable BB mass transfer. They estimate the eccentricity induced by the supernova kick in this formation scenario to be between $10^{-6}$ and $10^{-3}$ at 10 Hz. To measure the eccentricity of GW190425, \cite{10.1093/mnrasl/slaa084} reweight the posterior samples from the parameter estimation performed using circular binaries to estimate the limit of the eccentricity using the same method used to estimate the eccentricity of binary black holes \citep{Romero-Shaw:2019itr}. They estimate the eccentricity of GW190425 at 10 Hz to be $e \leq 0.007$ (90\% confidence) using a prior uniform in $\log e$. They find no evidence for or against unstable BB mass transfer as their analysis is not able to distinguish the small residual eccentricity expected from the investigated formation channel. 

\begin{figure*}
    \includegraphics[width=0.99\textwidth]{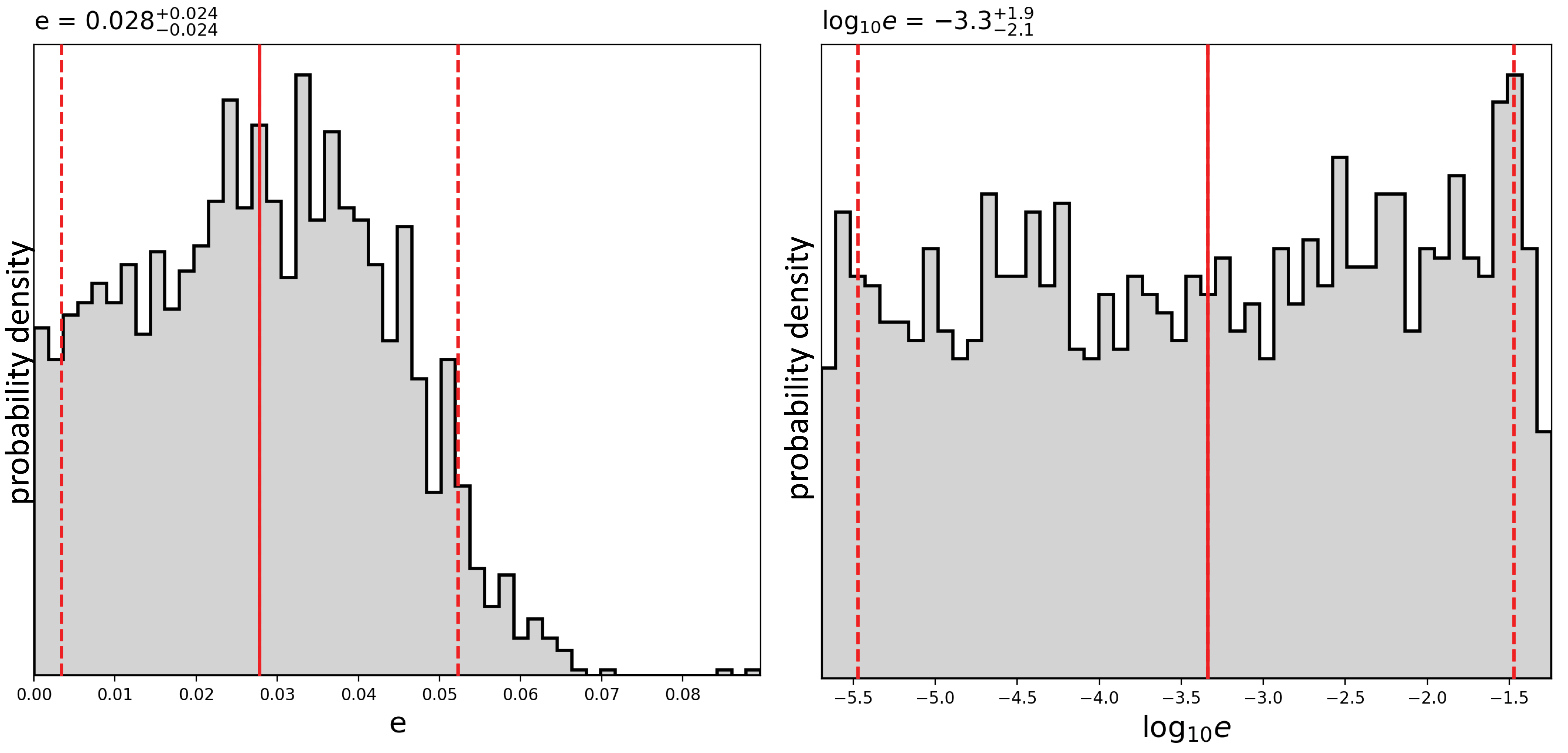}
    \caption{Eccentricity posteriors of GW190425 (solid black line) plotted against their priors (dotted line) for two choices of prior: uniform in $\ecc$ (left) and uniform in $\log_{10}(e)$ (right). We quote the median (solid red line) and 90\% credible interval (dashed red lines) for e in each posterior.}
    \label{Fig:1DMarginal}
\end{figure*}

To more directly compare our limit on GW190425's eccentricity, we repeat our analysis using a $\log e$ prior. In Figure~\ref{Fig:1DMarginal} we can see the differences in the posterior distributions of each prior. With the $\log e$ prior we estimate the eccentricity at 10 Hz to be $e \leq 0.023$. This is a factor of three larger than interval estimated by \cite{10.1093/mnrasl/slaa084}. By re-weighting the posterior samples rather than a full MCMC, the degeneracy between $\chirpm$ and $\ecc$ is missed. We find that by excluding posterior samples with lower values of $\chirpm$, we can recover the upper limit reported by \cite{10.1093/mnrasl/slaa084}. Although our limit on the eccentricity is larger than that of \cite{10.1093/mnrasl/slaa084}, our result does not change their conclusion: indeed the strong dependence of the eccentricity posterior on the prior seen in Figure~\ref{Fig:1DMarginal} agrees with their conclusion that the signal-to-noise ratio of GW190425 is not large enough to explore the eccentricities expected in BB mass transfer. We would need to be able to determine the eccentricity at lower frequencies to distinguish the formation channel.

\section{Conclusion}
Our analysis used the gravitational-wave observations as well as a prior on the eccentricity to constrain the eccentricity of GW170817 and GW190425. Our 90\% confidence limit using a uniform prior on $\ecc$ for GW170817, $e \leq 0.024$, and GW190425, $e \leq 0.048$, are consistent with expectations since they were found by a circular search \citep{Peters:1964zz}. We have constrained the eccentricity to a factor of two smaller than estimates obtained from circular searches \citep{Brown:2009ng,Huerta:2013qb}. Our 90\% credible intervals on the eccentricity of GW190425 are a approximately a factor of six larger than the interval estimated by \cite{10.1093/mnrasl/slaa084}, which used a prior uniform in $\log e$. This demonstrates the impact of prior choice, and the importance of measuring the eccentricity of signals using full parameter estimation to account for the correlation between parameters.

Unfortunately, based on current merger rate estimates the detection of an eccentric binary neutron star merger will be difficult with current observatories \citep{Lee:2009ca,Ye:2019xvf,Nitz:2019spj}, but gravitational-wave capture binaries that have $e \geq 0.8$ and could form in the LIGO-Virgo band \citep{Rodriguez:2018pss,Takatsy:2018euo}. However, since the eccentricity of the detections is expected to be low and negligible, $e \leq 0.02$, a circular search is effective in detecting them \citep{Brown:2009ng,Huerta:2013qb}. The detection of a binary neutron star mergers with high eccentricity or spin in future observing runs or with third-generation detectors \citep{Maggiore:2019uih,Reitze:2019iox} will reveal more about the formation channel of eccentric binaries and the existence of a dynamical formation channel.

\section{Data availability}
The data underlying this article are available in the associated data release on GitHub, at https://github.com/gwastro/bns-eccentric-pe ~\citep{bns-ecc-pe-release}.

\section*{Acknowledgements}

We acknowledge the Max Planck Gesellschaft for support and the Atlas cluster computing team at AEI Hannover. This research was supported in part by the National Science Foundation under Grant No.~PHY-1748958. DAB thanks National Science Foundation Grant No.~PHY-1707954 for support. AL thanks National Science Foundation Grant No.~AST-1559694 for support. This research has made use of data, software and/or web tools obtained from the Gravitational Wave Open Science Center (https://www.gw-openscience.org), a service of LIGO Laboratory, the LIGO Scientific Collaboration and the Virgo Collaboration. LIGO is funded by the U.S. National Science Foundation. Virgo is funded by the French Centre National de Recherche Scientifique (CNRS), the Italian Istituto Nazionale della Fisica Nucleare (INFN) and the Dutch Nikhef, with contributions by Polish and Hungarian institutes.\newline\newline\newline



\bibliographystyle{mnras}
\bibliography{references} 








\bsp	
\label{lastpage}
\end{document}